\documentstyle[prd,aps]{revtex}
\tighten
\begin{document}
\preprint{LPTHE-96/13; DEMIRM-96; astro-ph/9609129\bigskip}
\draft
\title{\bf  FRACTAL DIMENSIONS AND SCALING LAWS IN THE INTERSTELLAR
MEDIUM:  A NEW FIELD THEORY APPROACH}
\author{{\bf  H. J. de Vega$^{(a)}$,
  N. S\'anchez$^{(b)}$ and  F. Combes$^{(b)}$,}\bigskip}
\address
{ (a)  Laboratoire de Physique Th\'eorique et Hautes Energies,
Universit\'e Paris VI, Tour 16, 1er \'etage, 4, Place Jussieu
75252 Paris, Cedex 05, FRANCE. Laboratoire Associ\'e au CNRS UA 280.\\
(b) Observatoire de Paris,  Demirm, 61, Avenue de l'Observatoire,
75014 Paris,  FRANCE. 
Laboratoire Associ\'e au CNRS UA 336, Observatoire de Paris et
\'Ecole Normale Sup\'erieure.  \\ }
\date{astro-ph/9609129. {\bf To appear in Phys. Rev. D, November 15, 1996}}
\maketitle
\begin{abstract}
We develop a field theoretical approach to the cold interstellar medium (ISM). 
We show that a non-relativistic
self-gravitating gas in thermal equilibrium with variable
number of atoms or fragments is exactly equivalent to a field theory
of a single scalar field $ \phi({\vec x}) $ with exponential self-interaction.
We analyze this field theory perturbatively and non-perturbatively
through the renormalization group approach. We show {\bf scaling}  
behaviour (critical) for a continuous range of the temperature and of
the other  
physical parameters. We derive in this framework the scaling relation 
$ \Delta M(R) \sim R^{d_H} $ for the mass on a region of size $ R $,
and $ \Delta v \sim R^q $ for the velocity dispersion where 
$ q = \frac12(d_H -1) $. For the density-density correlations
we find a power-law  behaviour for large distances 
$ \sim |{\vec r_1} -{\vec r_2}|^{2 d_H -6} $.  The fractal dimension
$  d_H $ turns to be related with the critical exponent $ \nu $ of the 
correlation length by $  d_H = 1/ \nu $. The renormalization group
approach for a single component scalar field in three dimensions
states that the long-distance critical behaviour is governed by the
(non-perturbative) Ising fixed point. The corresponding 
values of the scaling exponents are  $  \nu = 0.631...  , \; d_H = 1.585... $
and $ q = 0.293...$. Mean field theory yields for the scaling
exponents $ \nu = 1/2  , \; d_H = 2 $ and $ q = 1/2 $. Both the Ising
and the mean field values are compatible with the present ISM
observational data: $  1.4    \leq   d_H    \leq   2     ,   \;
0.3  \leq     q  \leq 0.6 \;  $.

As typical in critical phenomena, the scaling behaviour and critical
exponents of the ISM can be obtained without dwelling into the dynamical
(time-dependent) behaviour. 

The relevant r\^ole of selfgravity is stressed by the authors in a
{\it Letter to Nature}, September 5, 1996.
\end{abstract}
\pacs{98.38.-j, 11.10.Hi, 05.70.Jk}

\section{Introduction and results}

The interstellar medium (ISM) is a gas essentially formed by atomic (HI) 
and molecular ($H_2$) hydrogen, distributed in cold ($T \sim 5-50 K$) 
clouds, in a very inhomogeneous and fragmented structure. 
These clouds are confined in the galactic plane 
and in particular along the spiral arms. They are distributed in 
a hierarchy of structures, of observed masses from 
$1\; M_{\odot}$ to $10^6 M_{\odot}$. The morphology and
kinematics of these structures are traced by radio astronomical 
observations of the HI hyperfine line at the wavelength of 21cm, and of
the rotational lines of the CO molecule (the fundamental line being
at 2.6mm in wavelength), and many other less abundant molecules.
  Structures have been measured directly in emission from
0.01pc to 100pc, and there is some evidence in VLBI (very long based 
interferometry) HI absorption of structures as low as $10^{-4}\; pc = 20$ AU 
(3 $10^{14}\; cm$). The mean density of structures is roughly inversely
proportional to their sizes, and vary 
between $10$ and $10^{5} \; atoms/cm^3$ (significantly above the 
mean density of the ISM which is about 
$0.1 \; atoms/cm^3$ or $1.6 \; 10^{-25}\; g/cm^3$ ).
Observations of the ISM revealed remarkable relations between the mass, 
the radius and velocity dispersion of the various regions, as first 
noticed by Larson \cite{larson}, and  since then confirmed by many other 
independent observations (see for example ref.\cite{obser}). 
From a compilation of well established samples of data for many different  
types of molecular clouds of maximum linear dimension (size) $ R $,  
 mass fluctuation $ \Delta M$ and internal velocity dispersion $
\Delta v$ in each region:  
\begin{equation}\label{vobser}
\Delta M (R)  \sim    R^{d_H}     \quad        ,     \quad  \Delta v
\sim R^q \; , 
\end{equation}
over a large range of cloud sizes, with   $ 10^{-4}\; - \; 10^{-2}
\;  pc \;   \leq     R   \leq 100\;  pc, \;$
\begin{equation}\label{expos}
1.4    \leq   d_H    \leq   2     ,   \;     0.3  \leq     q  \leq
0.6 \; . 
\end{equation}
These {\bf scaling}  relations indicate a hierarchical structure for the 
molecular clouds which is independent of the scale over the above 
cited range; above $100$ pc in size, corresponding to giant molecular clouds,
larger structures will be destroyed by galactic shear.

These relations appear to be {\bf universal}, the exponents 
$d_H , \; q$ are almost constant over all scales of the Galaxy, and
whatever be  
the observed molecule or element. These properties of interstellar cold 
gas are supported first at all from observations (and for many different 
tracers of cloud structures: dark globules using $^{13}$CO, since the
more abundant isotopic species $^{12}$CO is highly optically thick, 
dark cloud cores using $HCN$ or $CS$ as density tracers,
 giant molecular clouds using $^{12}$CO, HI to trace more diffuse gas, 
and even cold dust emission in the far-infrared).
Nearby molecular clouds are observed to be fragmented and 
self-similar in projection over a range of scales and densities of 
at least $10^4$, and perhaps up to $10^6$.

The physical origin as well as the interpretation of the scaling relations 
 (\ref{vobser}) are not theoretically understood. The theoretical
 derivation of these 
 relations has been the subject of many proposals and controversial 
discussions. It is not our aim here to account for all the proposed models 
of the ISM and we refer the reader to refs.\cite{obser} for a review.

The physics of the ISM is complex, especially when we consider the violent
perturbations brought by star formation. Energy is then poured into 
the ISM either mechanically through supernovae explosions, stellar winds,
bipolar gas flows, etc.. or radiatively through star light, heating or
ionising the medium, directly or through heated dust. Relative velocities
between the various fragments of the ISM exceed their internal thermal
speeds, shock fronts develop and are highly dissipative; radiative cooling
is very efficient, so that globally the ISM might be considered 
isothermal on large-scales. 
Whatever the diversity of the processes, the universality of the
scaling relations suggests a common mechanism underlying the physics.
  We propose that self-gravity is the main force at the origin of the 
structures, that can be perturbed locally by heating sources. 
Observations are compatible with virialised structures at all scales.
 Moreover, it has been suggested that the molecular clouds ensemble is
in isothermal equilibrium with the cosmic background radiation at $T \sim 3 K$
in the outer parts of galaxies, devoid of any star and heating
sources \cite{pcm}. This colder isothermal medium might represent the ideal
frame to understand the role of self-gravity in shaping the hierarchical
structures. Our aim is to show that the scaling laws obtained are then
quite stable to perturbations.

Till now, no theoretical derivation of the scaling laws
eq.(\ref{vobser}) has been  
provided in which the values of the exponents are {\bf obtained}
  from the theory
 (and not just taken from outside or as a starting input or hypothesis).

The aim of these authors is to develop a theory of the cold ISM. A first 
step in this goal is to provide a theoretical derivation of the scaling 
laws eq.(\ref{vobser}), in which the values of the exponents $d_H , \; q$ are
{\bf obtained} from the theory. 
For this purpose, we will implement for the ISM the powerful tool of field 
theory and the Wilson's approach to critical phenomena \cite{kgw}.
 
We  consider a gas of non-relativistic atoms interacting with each other
 through Newtonian gravity and which are in thermal
equilibrium at temperature $ T $.
We  work in the grand canonical ensemble, allowing for a variable
number of particles $N$.

Then, we show that this system 
is exactly equivalent to a field theory of a single scalar field
$\phi({\vec x})$ with 
exponential  interaction. We express the grand canonical partition function
$ {\cal Z} $ as
\begin{equation}\label{zetafi}
{\cal Z} =  \int\int\;  {\cal D}\phi\;  e^{-S[\phi(.)]} \; , 
\end{equation}
where
\begin{eqnarray}\label{SmuyT}
S[\phi(.)] & \equiv &  {1\over{T_{eff}}}\;
\int d^3x \left[ \frac12(\nabla\phi)^2 \; - \mu^2 \; 
e^{\phi({\vec x})}\right] \; , \cr \cr
 T_{eff} &=& 4\pi \; {{G\; m^2}\over {T}} \quad , \quad
\mu^2 = \sqrt{2\over {\pi}}\; z\; G \, m^{7/2} \, \sqrt{T} \; ,
\end{eqnarray}
$ m $ stands for the mass of the atoms and $ z $  for the fugacity. 
We show that in the $\phi$-field language, the particle density 
expresses as
\begin{equation}\label{denfi}
  <\rho({\vec r})> =  -{1 \over {T_{eff}}}\;<\nabla^2 \phi({\vec r})>=
{{\mu^2}\over{T_{eff}}} \; <e^{\phi({\vec r})}>  \; .
\end{equation}
where $ <\ldots > $ means functional average over   $ \phi(.) $
with statistical weight $  e^{-S[\phi(.)]} $. Density correlators are
written as
\begin{eqnarray}\label{correI}
C({\vec r_1},{\vec r_2})&\equiv&
<\rho({\vec r_1})\rho({\vec r_2}) > -<\rho({\vec r_1})><\rho({\vec r_2}) > 
\cr \cr
&=&  {{\mu^4}\over{T_{eff}}^2} \; \left[  
<e^{\phi({\vec r_1})} \; e^{\phi({\vec r_2})}> -
<e^{\phi({\vec r_1})}> \; <e^{\phi({\vec r_2})}> \right]\; .
\end{eqnarray}
The  $\phi$-field defined by eqs.(\ref{zetafi})-(\ref{SmuyT}) has remarkable
properties under  scale transformations 
$$
{\vec x} \to {\vec x}_{\lambda} \equiv \lambda{\vec x} \; ,
$$
where $\lambda$ is an arbitrary real number. For any solution  $
\phi({\vec x}) $ of the stationary point equations,
\begin{equation}\label{eqMovI}
\nabla^2\phi({\vec x}) +  \mu^2 \; e^{\phi({\vec x})} = 0 \; ,
\end{equation}
there is a family of dilated solutions of the same equation (\ref{eqMovI}),
given by
$$
\phi_{\lambda}({\vec x}) \equiv \phi(\lambda{\vec x}) +\log\lambda^2
\; .
$$
In addition, $ S[\phi_{\lambda}(.)] = \lambda^{2-D} \; S[\phi(.)] $. 

\bigskip

We study the field theory (\ref{zetafi})-(\ref{SmuyT}) both
perturbatively and non-perturbatively.

The computation of the thermal fluctuations through the evaluation of
the functional integral  eq.(\ref{zetafi}) is quite non-trivial. We
use the scaling property as a guiding principle. In order to built a
perturbation theory in the dimensionless coupling  $ g \equiv
\sqrt{\mu \, T_{eff}} $ we look for stationary points of
eq.(\ref{SmuyT}). We compute the density correlator eq.(\ref{correI}) to
leading order in $ g $. For large distances it behaves as

\begin{equation}\label{coRa}
 C({\vec r_1},{\vec r_2}) \buildrel{  | {\vec  r_1} - {\vec  r_2}|\to
\infty}\over =  {{ \mu^4 }\over {32\, \pi^2 \; 
 | {\vec  r_1} - {\vec  r_2}|^2}} + O\left( \; | {\vec
r_1} - {\vec r_2}|^{-3}\right)\; . 
\end{equation}

We  analyze further this theory with the renormalization group
approach. Such non-perturbative approach is the more powerful
framework to
derive scaling behaviours in field theory \cite{kgw,dg,nn}.

We show that the mass contained in a region of volume $ V =R^3 $ scales as
$$
<M(R)> =    m   \; \int^R
<e^{\phi({\vec x})}> \; d^3x   \simeq  m  \, V \, a +m  \, {K
\over{1-\alpha}}\; R^{ \frac1{\nu}} + \ldots\; , 
$$
and the mass fluctuation, $ (\Delta M(R))^2 = <M^2>-<M>^2 $,
scales as
$$
\Delta M(R)  \sim  R^{d_H}\; .
$$
Here $ \nu $ is the correlation length critical exponent for the
$\phi$-theory (\ref{zetafi}) and $ a $ and $ K $ are  constants. Moreover, 
\begin{equation}\label{SdensiI}
<\rho({\vec r})> = m  a \; + m \, {K
\over{4\pi\nu(1-\alpha)}}\;r^{ \frac1{\nu}-3} \quad {\rm for}\; r
\;  {\rm of ~ order}\;\sim R \; .
\end{equation}
The scaling exponent $ \nu $ can be identified with the inverse
Haussdorf (fractal) dimension $d_H$ of the system
$$
d_H = \frac1{\nu} \; .
$$
In this way, $ \Delta M\sim R^{d_H} $ according to the usual
definition of fractal dimensions \cite{sta}.


From the renormalization group analysis, 
the density-density correlators (\ref{correI}) result to be, 
\begin{equation}\label{corI}
C({\vec r_1},{\vec r_2})\sim |{\vec r_1} -{\vec r_2}|^{\frac2{\nu} -6} \; .
\end{equation}
Computing the average gravitational potential energy and using the
virial theorem yields for the velocity dispersion,
$$
\Delta v \sim R^{\frac12(\frac1{\nu} -1)} \; .
$$
This gives a new scaling relation between the exponents $ d_H $ and $ q $
$$
q =\frac12\left(\frac1{\nu} -1\right) =\frac12(d_H -1)  \; .
$$

The perturbative calculation (\ref{coRa}) yields the mean field value
for $ \nu $ \cite{ll}. That is,
\begin{equation}\label{meanF}
\nu= \frac12  \quad ,  \quad d_H = 2 \quad {\rm and } \quad q = \frac12 \; .
\end{equation}
 
We find scaling behaviour in the $\phi$-theory for a {\bf continuum set} of
values of $\mu^2$ and $ T_{eff} $.
The renormalization group transformation amounts to replace 
the parameters $ \mu^2 $ and $ T_{eff} $ 
in $ \beta\, H $ and $ S[\phi(.)] $ by the effective ones at the scale
$ L $ in question. 

The renormalization group approach applied to a
{\bf single}  component scalar field in three space dimensions
indicates that the long distance critical behaviour is governed by the
(non-perturbative) Ising fixed point  \cite{kgw,dg,nn}. 
Very probably, there are no further fixed points \cite{grexa}. 
The scaling exponents associated to the Ising fixed point are

\begin{equation}\label{Isint}
\nu = 0.631...  \quad , \quad d_H = 1.585...   \quad {\rm and} \quad
 q = 0.293...\; \; .
\end{equation}

Both the mean field  (\ref{meanF}) and the Ising  (\ref{Isint})
numerical values are compatible
with the present observational values  (\ref{vobser}) - (\ref{expos}).

\bigskip

The theory presented here also predicts a power-law behaviour for
the two-points ISM density correlation function (see eq.(\ref{corI}),
$ 2 d_H - 6 = - 2.830\ldots$, for the Ising fixed point
and $ 2 d_H - 6 = - 2 $ for the mean field exponents),
that should be compared with observations. Previous attempts to
derive correlation functions from observations were not entirely conclusive, 
because of lack of dynamical range \cite{klein}, but much more extended maps of
the ISM could be available soon to test our theory. In addition, we predict 
an independent exponent for the gravitational 
potential correlations ($ \sim r^{-1-\eta} $, where
$ \eta_{Ising}=0.037\ldots $ and $ \eta_{mean ~ field} = 0 $
\cite{dg}), which could be checked through 
gravitational lenses observations in front of quasars.

\bigskip

The mass parameter $\mu $ [see eq.(\ref{SmuyT})] in the $\phi$-theory
turns to coincide at the tree level with the inverse of the Jeans length
$$
\mu =  \sqrt{12 \over {\pi}}\; { 1 \over {d_J}} \; .
$$
We find that in the scaling domain  the Jeans distance $ d_J $ grows
as $  <d_J> \sim R $. This shows that the Jeans distance   {\bf scales} 
with the  {\bf size} of the system and therefore the instability is
 present for all sizes $ R $. Had  $ d_J $ being  of order larger than
$ R $, the Jeans instability would be absent. 

\bigskip

The gravitational gas in thermal equilibrium explains quantitatively
the observed scaling laws in the ISM. This fact does not exclude  
turbulent phenomena in the ISM. 
Fluid flows (including turbulent regimes) are probably relevant
in the dynamics (time dependent processes) of the ISM. As usual in critical
phenomena \cite{kgw,dg}, the equilibrium scaling laws  can be  understood 
for the ISM without  dwelling with the dynamics. 
A further step in the study of the ISM will be to include the
dynamical (time dependent) description within the field theory
approach presented in this paper.

\bigskip

If the ISM is considered as a flow,  the Reynolds number $Re_{ISM}$ on
scales  $L \sim 100$pc  has a very high value of the order of $10^6$.  
This led to the suggestion that the ISM (and the universe in general)
could be {\bf modelled}  as a turbulent flow \cite{weisz}. 
(Larson \cite{larson} first observed that the 
exponent in the power-law relation for the velocity dispersion is not greatly 
different from the Kolmogorov value $1/3$ for subsonic turbulence).

It must be noticed that the turbulence hypothesis for the ISM is based on 
the comparison of the ISM with the results known for incompressible 
flows. However, the physical conditions in the ISM are 
very different from those of incompressible flows in the laboratory. 
(And the 
study of ISM turbulence needs more complete and enlarged investigation 
than those performed until now based in the concepts of flow turbulence 
in the laboratory).  
Besides the facts that the ISM exhibits large density fluctuations on all 
scales, and the observed fluctuations are highly supersonic, (thus the 
ISM can not viewed as an `incompressible' and  `subsonic' flow),
 and besides other differences, an essential feature to point out is that
 the long-range self-gravity interaction present in the ISM is completely 
absent in the studies of flow turbulence. 
In any case, in a satisfactory theory of the ISM, 
it should be possible to extract the  behaviours of 
the ISM (be turbulent or whatever) from the theory 
as a result, instead to be introduced as a starting input or  hypothesis.

This paper is organized as follows. In section II we develop the field
theory approach to the gravitational gas. A short distance cutoff is
naturally present here and prevents zero distance gravitational
collapse singularities (which would be unphysical in the present
case). Here, the cutoff theory is physically meaningful. The
gravitational gas is also treated in a $D$-dimensional space.

In section III we study the scaling behaviour and thermal fluctuations
both in perturbation theory and non-perturbatively (renormalization
group approach). $g^2 \equiv \mu\, T_{eff} $ acts as the dimensionless
coupling constant for the non-linear fluctuations of the field $\phi$.
We show that these fluctuations are massless and that the theory
scales (behaves critically) for a continuous range of values  $ \mu^2
\; T_{eff} $. Thus, changing $ \mu^2 $ and $ T_{eff} $ keeps the
theory at {\bf criticality}. The renormalization group analysis made
in section III confirm such results. We also treat (sect. III.E) the
two dimensional case making contact with random surfaces and their
fractal dimensions. 

Discussion and remarks are presented in section IV. External gravity forces to
the gas like stars are shown {\bf not} to affect the scaling behaviour
of the gas. That is, the scaling exponents $ q , \; d_H $ are solely
governed by fixed points and hence, they are stable under  gravitational
perturbations.
In addition, we generalize the $\phi$-theory  to a gas formed by
several types of atoms with different masses and fugacities. Again, the 
scaling exponents are shown to be identical to the gravitational gas 
formed of identical atoms.

The differences between the critical
behaviour of the gravitational gas and those in spin models (and other
statistical models in the same universality class) are also pointed
out in sec. IV.

\section{Field theory approach to the gravitational gas}

Let us  consider a gas of non-relativistic atoms with mass $m$ interacting
only through Newtonian gravity and which are in thermal
equilibrium at temperature $ T \equiv \beta^{-1} $.
We shall work in the grand canonical ensemble, allowing for a variable
number of particles $N$.

The grand partition function of the system can be written as

\begin{equation}\label{gfp}
{\cal Z} = \sum_{N=0}^{\infty}\; {{z^N}\over{N!}}\; \int\ldots \int
\prod_{l=1}^N\;{{d^3p_l\, d^3q_l}\over{(2\pi)^3}}\; e^{- \beta H_N}
\end{equation}
where
\begin{equation}\label{hami3}
H_N = \sum_{l=1}^N\;{{p_l^2}\over{2m}} - G \, m^2 \sum_{1\leq l < j\leq N}
{1 \over { |{\vec q}_l - {\vec q}_j|}}
\end{equation}
$G$ is Newton's constant and $z$ is the fugacity.

The integrals over the momenta $p_l, \; (1 \leq l \leq N) $
can be performed explicitly in eq.(\ref{gfp}) 
using
$$
\int\;{{d^3p}\over{(2\pi)^3}}\; e^{- {{\beta p^2}\over{2m}}} =
\left({m \over{2\pi \beta}}\right)^{3/2}
$$
We thus find,
\begin{equation}\label{gfp2}
\displaystyle{
{\cal Z} = \sum_{N=0}^{\infty}\; {1 \over{N!}}\;
\left [ z\left({m \over{2\pi \beta}}\right)^{3/2}\right]^N
\; \int\ldots \int
\prod_{l=1}^N d^3q_l\;\; e^{ \beta G \, m^2 \sum_{1\leq l < j\leq N}
{1 \over { |{\vec q}_l - {\vec q}_j|}} }}
\end{equation}

We proceed now to recast this many-body problem into a field theoretical
form \cite{origen,stra,sam,kh}.
 
Let us define the density
\begin{equation}\label{defro}
\rho({\vec r})= \sum_{j=1}^N\; \delta({\vec r}- {\vec q}_j)\; ,
\end{equation}
such that, we can rewrite the potential energy in eq.(\ref{gfp2}) as
\begin{equation}\label{PotE}
 \frac12 \, \beta G \, m^2 \sum_{1\leq l \neq j\leq N}
{1 \over { |{\vec q}_l - {\vec q}_j|}} =  \frac12\,  \beta \, G \, m^2
\int_{ | {\vec x} - {\vec y}|> a}\;
{{d^3x\, d^3y}\over { | {\vec x} - {\vec y}|}}\; \rho({\vec x})
\rho({\vec y}) \; .
\end{equation}
The cutoff $ a $ in the r.h.s. is introduced in order to avoid
self-interacting divergent terms. However, such divergent terms would
contribute to ${\cal Z}$ by
an infinite multiplicative factor that can be factored out.

By using
$$ 
\nabla^2 { 1 \over { | {\vec x} - {\vec y}|}}= -4\pi \; \delta( {\vec
x} - {\vec y}) \; ,
$$
and partial integration we can now represent the exponent of the
potential energy eq.(\ref{PotE}) as a functional integral\cite{stra}
\begin{equation}\label{reprf}
e^{  \frac12\, \beta G \, m^2
\int \;
{{d^3x\, d^3y}\over { | {\vec x} - {\vec y}|}}\; \rho({\vec x})
\rho({\vec y})} = \int\int\; {\cal D}\xi \; e^{ -\frac12\int d^3x \; (\nabla
\xi)^2 \; + \; 2 m \sqrt{\pi G\beta}\; \int d^3x \; \xi({\vec x})\;
\rho({\vec x}) } 
\end{equation}

Inserting this expression into eq.(\ref{gfp2})  and using
eq.(\ref{defro}) yields 
\begin{eqnarray}\label{gfp3} 
{\cal Z} &=& \sum_{N=0}^{\infty}\; {1 \over{N!}}\;
\left [ z\left({m \over{2\pi \beta}}\right)^{3/2}\right]^N\;
 \int\int\;  {\cal D}\xi \; e^{ -\frac12\int d^3x \; (\nabla \xi)^2}
\; \int\ldots \int
\prod_{l=1}^N d^3q_l\; \; e^{ 2 m  \sqrt{\pi G\beta}\; \sum_{l=1}^N
\xi({\vec q}_l)} \cr \cr
 &=& \int\int\;  {\cal D}\xi \; e^{ -\frac12\int d^3x  \;(\nabla \xi)^2}\;
 \sum_{N=0}^{\infty}\; {1 \over{N!}}\;
\left [ z\left({m \over{2\pi \beta}}\right)^{3/2}\right]^N\;
\left[ \int d^3q \;  e^{ 2 m  \sqrt{\pi G\beta}\;\xi({\vec q})}
\right]^N \cr \cr
 &=& \int\int\;  {\cal D}\xi \; e^{ -\int d^3x \left[ \frac12(\nabla \xi)^2\;
- z \left({m \over{2\pi \beta}}\right)^{3/2}\; e^{ 2 m \sqrt{\pi
G\beta}\;\xi({\vec x})}\right]} \; \quad .
\end{eqnarray}

It is convenient to introduce the dimensionless field
\begin{equation}
\phi({\vec x}) \equiv  2 m \sqrt{\pi G\beta}\;\xi({\vec x}) \; .
\end{equation}

Then,
\begin{equation}\label{zfi}
{\cal Z} =  \int\int\;  {\cal D}\phi\;  e^{ -{1\over{T_{eff}}}\;
\int d^3x \left[ \frac12(\nabla\phi)^2 \; - \mu^2 \; e^{\phi({\vec
x})}\right]}\; , 
\end{equation}

where

\begin{equation}\label{muyT}
\mu^2 = \sqrt{2\over {\pi}}\; z\; G \, m^{7/2} \, \sqrt{T} 
\quad , \quad T_{eff} = 4\pi \; {{G\; m^2}\over {T}} \; .
\end{equation}
The partition function for the gas of particles in gravitational
interaction has been transformed into the partition function for a single
scalar field $\phi({\vec x})$  with  {\bf local} action
\begin{equation}\label{acci}
S[\phi(.)] \equiv  {1\over{T_{eff}}}\;
\int d^3x \left[ \frac12(\nabla\phi)^2 \; - \mu^2 \; e^{\phi({\vec
x})}\right] \; .
\end{equation}
The  $\phi$  field exhibits an exponential self-interaction $ - \mu^2
\; e^{\phi({\vec x})} $. 

Notice that the effective 
temperature $ T_{eff} $ for the  $\phi$-field partition function 
turns out to be  {\bf inversely}
proportional to $ T $ whereas the characteristic length $\mu^{-1}$ behaves
as $ \sim T ^{-1/4}$. This is a duality-type mapping between the two models.

It must be noticed that the term $ - \mu^2 \; e^{\phi({\vec x})} $
makes the  $\phi$-field energy density unbounded from
below. Actually, the initial Hamiltonian (\ref{gfp}) is also  unbounded from
below. This unboundness physically originates in the attractive
character of the gravitational force. Including a short-distance
cutoff [see sec. 2A, below] eliminates the zero distance singularity
and hence the possibility  of zero-distance
collapse which is unphysical in the present context. 
We therefore expect meaningful physical results in the
cutoff theory. Moreover, assuming zero boundary conditions for
$\phi({\vec r})$ at $ r \to \infty $ shows that the derivatives of
$\phi$ must also be large if $ e^\phi$ is large. Hence, the term $
\frac12(\nabla\phi)^2 $ may stabilize the energy.

The action  (\ref{acci}) defines a non-renormalizable field
theory for any number of dimensions $ D > 2 $ [see
eq.(\ref{zfiD})]. This is a further reason to keep the short-distance
cutoff non-zero.

\bigskip

Let us compute now the statistical average  value of the density
$\rho({\vec r})$ which in  the grand canonical ensemble is given by
\begin{equation}
<\rho({\vec r})> =   {\cal Z}^{-1}\; \sum_{N=0}^{\infty}\; {1 \over{N!}}\;
\left [ z\left({m \over{2\pi \beta}}\right)^{3/2}\right]^N
\; \int\ldots \int
\prod_{l=1}^N d^3q_l\; \; \rho({\vec r}) \; 
e^{ \frac12\, \beta G \, m^2 \sum_{1\leq l \neq j\leq N}
{1 \over { |{\vec q}_l - {\vec q}_j|}} }\; .
\end{equation}

As usual in the functional integral calculations, 
it is convenient to introduce sources in the partition function (\ref{zfi})
in order to compute  average values of fields

\begin{equation}\label{zfiJ}
{\cal Z}[J(.)] \equiv  \int\int\; {\cal D}\phi\;  e^{ -{1\over{T_{eff}}}\;
\int d^3x \left[ \frac12(\nabla
\phi)^2 \; - \mu^2 \; e^{\phi({\vec x})}\; \right]
+\int d^3x  \;J({\vec x})\; \phi({\vec x}) \; }\; .
\end{equation}
The average value of $ \phi({\vec r}) $ then writes as
\begin{equation}
< \phi({\vec r})> = {{\delta \log{\cal Z} }\over{\delta J({\vec r})}}\; .
\end{equation}

In order to compute $<\rho({\vec r})>$ it is useful to introduce
\begin{equation}
{\cal V}[J(.)] \equiv  \frac12 \,\beta G \, m^2
\int_{ | {\vec x} - {\vec y}|> a }\;
{{d^3x\, d^3y}\over { | {\vec x} - {\vec y}|}}\; 
\left[ \rho({\vec x})+ \;J({\vec x})\;\right]
\left[\rho({\vec y})+ \;J({\vec y})\;\right]\; .
\end{equation}
Then, we have
$$
 \rho({\vec r}) \; e^{{\cal V}[0]} = -{1 \over{T_{eff}}} \; \nabla^2_{\vec
r} \left({{\delta}\over{\delta J({\vec r})}} e^{{\cal V}[J(.)]}
\right)|_{J=0}\; .
$$

By following the same steps as in eqs.(\ref{reprf})-(\ref{gfp3}), we find
\begin{eqnarray}
<\rho({\vec r})> &=&  -{1 \over{T_{eff}}} \; \nabla^2_{\vec
r} \left({{\delta}\over{\delta J({\vec r})}}
  \sum_{N=0}^{\infty}\; {1 \over{N!}}\;
\left [ z\left({m \over{2\pi \beta}}\right)^{3/2}\right]^N
\; \;   {\cal Z}[0]^{-1} \right.\cr \cr
\int\int   \;   {\cal D}\xi & & \left. e^{ -\int d^3x\left[\frac12 \;
(\nabla \xi)^2 
- 2 m  \sqrt{\pi G\beta}\;\xi({\vec x})\;  J({\vec x})\right]}\;
\; \int\ldots \int
\prod_{l=1}^N d^3q_l\; \; e^{ 2 m  \sqrt{\pi G\beta}\; \sum_{l=1}^N
\xi({\vec q}_l)}\right)|_{J=0} \cr\cr
&=&  -{1 \over{T_{eff}}} \; \nabla^2_{\vec
r} \left({{\delta}\over{\delta J({\vec r})}}\;\log {\cal Z}[J(.)]\right)|_{J=0}
 \quad .
\end{eqnarray}

Performing the derivatives in the last formula yields
\begin{equation}
<\rho({\vec r})> = - {1 \over {T_{eff}}}\;   \int\int\; {\cal D}\phi\; \; 
\nabla^2 \phi({\vec r})\;
e^{-{1\over{T_{eff}}}\; \int d^3x \left[ \frac12(\nabla
\phi)^2 \; - \mu^2 \; e^{\phi({\vec x})}\;\right]}\; {\cal Z}[0]^{-1}\; .
\end{equation}
One can analogously prove that $ \rho({\vec r}) $ inserted in any
correlator becomes $ -{1 \over {T_{eff}}}\;  \nabla^2 \phi({\vec r}) $ 
in the $\phi$-field language. Therefore, we can express  the particle density
operator as
\begin{equation}\label{rouno}
 \rho({\vec r}) =  -{1 \over {T_{eff}}}\;  \nabla^2 \phi({\vec r}) \; .
\end{equation}

Let us now derive the field theoretical equations of motion. Since the
functional integral of a total functional derivative identically
vanishes, we can write
$$
  \int\int\; {\cal D}\phi\; \;\left[ - {{\delta S
}\over{\delta \phi({\vec r})}} + J({\vec r}) \right] e^{-S[\phi(.)] + 
\int d^3x  \;J({\vec x})\; \phi({\vec x}) \; } = 0
$$
We get from eq.(\ref{acci})
$$
 {{\delta S}\over{\delta \phi({\vec r})}} = -  {1\over{T_{eff}}}\;
\left[ \nabla^2\phi({\vec r}) \; + \mu^2 \; e^{\phi({\vec
r})}\right]\; .
$$
Thus, setting $  J({\vec r}) \equiv 0 $,
\begin{equation}\label{ecmov}
< \nabla^2\phi({\vec r}) >  + \; \mu^2 \; <e^{\phi({\vec r})}> = 0
\end{equation}
Now, combining eqs.(\ref{rouno}) and (\ref{ecmov}) yields
\begin{equation}\label{densi}
  <\rho({\vec r})>={{\mu^2}\over{T_{eff}}} \; <e^{\phi({\vec r})}>  \; . 
\end{equation}

\bigskip

By using eq.(\ref{rouno}), the gravitational potential at the point $ \vec r $ 
$$
U( \vec r ) = -G m \int {{d^3x} \over  { | {\vec x} - {\vec r}|}}\; 
 \rho({\vec x}) \; ,
$$
can  be expressed as
\begin{equation}\label{Ufi}
U( \vec r ) = - {T \over m}\; \phi( \vec r ) \; .
\end{equation}

We can analogously express the correlation functions as
\begin{eqnarray}\label{corre}
C({\vec r_1},{\vec r_2})&\equiv&
<\rho({\vec r_1})\rho({\vec r_2}) > -<\rho({\vec r_1})><\rho({\vec r_2}) > 
\cr \cr
&=&  \left({1 \over{T_{eff}}} \right)^2\; \nabla^2_{\vec r_1}\;
\nabla^2_{\vec r_2}  \;
\left({{\delta}\over{\delta J({\vec r_1})}}\;{{\delta}\over{\delta
J({\vec r_2})}}\; \log{\cal Z}[J(.)]\right)|_{J=0} \; .
\end{eqnarray}
This can be also written as
\begin{equation}\label{corr2}
C({\vec r_1},{\vec r_2}) =  {{\mu^4}\over{T_{eff}}^2} \; \left[  
<e^{\phi({\vec r_1})} \; e^{\phi({\vec r_2})}> -
<e^{\phi({\vec r_1})}> \; <e^{\phi({\vec r_2})}> \right]\; .
\end{equation}

\subsection{Short distances cutoff}

A simple short distance regularization of the Newtonian force for the
two-body potential is
$$
v_a({\vec r}) = -{{G m^2} \over r}\; [ 1 - \theta(a-r) ] \; ,
$$
$ \theta(x)$ being the step function. The cutoff $ a $ can be chosen of
the order of atomic distances but its actual value is unessential.

The $N$-particle regularized Hamiltonian takes then the  form
\begin{equation}\label{hamiR}
H_N = \sum_{l=1}^N\;{{p_l^2}\over{2m}} + \frac12\,
 \sum_{1\leq l, j\leq N} \; v_a({\vec q}_l - {\vec q}_j) \; .
\end{equation}
Notice that now we can include in the sum terms with $l = j $ 
since $ v_a(0) = 0 $.

The steps from eq.(\ref{hami3}) to  eq.(\ref{zfi}) can be just
repeated by using now the regularized $v_a({\vec r})$. Notice that we
must use now the inverse operator of $ v_a({\vec r}) $ instead of that
of  $ 1/r , \; \left[ -\frac1{4\pi}\nabla^2 \right] $ , previously used.

We now find,
\begin{equation}\label{zfiR}
{\cal Z}_a =  \int\int\; {\cal D}\phi\;  e^{ -{1\over{T_{eff}}}\;
\int d^3x \left[ \frac12\phi K_a \phi \; - \mu^2 \; e^{\phi({\vec
x})}\right]}\; , 
\end{equation}
i. e.
\begin{equation}\label{Sregu}
S_a[\phi(.)] = {1\over{T_{eff}}}\;
\int d^3x \left[ \frac12\phi K_a \phi \; - \mu^2 \; e^{\phi({\vec
x})}\right] \; ,
\end{equation}
where $ K_a $ is the inverse operator of  $ v_a $,
\begin{eqnarray}
 K_a \phi ({\vec r}) &=& \int  K_a ({\vec r}-{\vec r}\, ') \;
\phi ({\vec r}\,')\; d^3r' \cr \cr 
\int  K_a({\vec r}-{\vec r}\,'')\; &{1 \over  {4\pi}}&\;
{ {1 - \theta(a- |{\vec r}\,''-{\vec r}\,'|
) }\over {|{\vec r}\,''-{\vec r}\,'|}}\;  d^3r'' =
\delta ({\vec r}-{\vec r}\,')\nonumber 
\end{eqnarray}
$ K_a ({\vec r})$ admits the Fourier representation,
$$
 K_a ({\vec r}) = V.P.\int {{d^3p}\over {(2\pi)^3}}\; {{p^2\; e^{i {\vec
p}.{\vec r}}}\over {\cos pa}}\; .
$$
Actually, $ K_a ({\vec r}) = 0 $ for $r \neq 0$. $ K_a ({\vec r})$ has
the following asymptotic expansion in powers of the cutoff $ a^2 $
\begin{equation}\label{deska}
 K_a ({\vec r}) = -\nabla^2 \delta ({\vec r}) + {{a^2}\over 2} \; 
\nabla^4 \delta ({\vec r}) + O(a^4) \; ,
\end{equation}
and then
\begin{equation}\label{Sar}
S_a[\phi(.)] = S[\phi(.)]
+ {{a^2}\over 2} \;  \int  d^3x\; (\nabla^2\phi)^2 + O(a^4) \; .
\end{equation}
As we see, the high orders in  $ a^2 $  are irrelevant operators
which do not affect  the scaling behaviour, as is well known from
renormalization group arguments. For $a \to 0$, the action (\ref{acci}) is
recovered.

\subsection{D-dimensional generalization}

This approach generalizes immediately to $D$-dimensional space where
the Hamiltonian (\ref{hami3}) takes then the form
\begin{equation}\label{hamiD}
H_N = \sum_{l=1}^N\;{{p_l^2}\over{2m}} - G \, m^2 \sum_{1\leq l < j\leq N}
{1 \over { |{\vec q}_l - {\vec q}_j|^{D-2}}},\quad  {\rm for}\;  D \neq 2
\end{equation}
and
\begin{equation}\label{hami2}
H_N = \sum_{l=1}^N\;{{p_l^2}\over{2m}} - G \, m^2 \sum_{1\leq l < j\leq N}
\log{1 \over { |{\vec q}_l - {\vec q}_j|}}, \quad  {\rm at}\;  D= 2\; .
\end{equation}

The steps from eq.(\ref{gfp}) to (\ref{zfi}) can be trivially
generalized with the help of the relation
\begin{equation}\label{Dgreen} 
\nabla^2 { 1 \over { | {\vec x} - {\vec y}|^{D-2}}}= -C_D \; \delta( {\vec
x} - {\vec y}) \; 
\end{equation}
in $D$-dimensions and
$$
\nabla^2 \log{ 1 \over { | {\vec x} - {\vec y}|}}= -C_2 \; \delta( {\vec
x} - {\vec y}) \; 
$$
at $ D= 2$. 

Here,
\begin{equation}
C_D \equiv (D-2)\, {{2 \pi^{D/2}}\over {\Gamma(\frac{D}{2})}}\; \; 
{\rm for}~~D\neq 2 \quad {\rm and}~~ C_2 \equiv 2\pi\; .
\end{equation}

We finally obtain as a generalization of  eq.(\ref{zfi}),
\begin{equation}\label{zfiD}
{\cal Z} =  \int\int\;  {\cal D}\phi\;  e^{ -{1\over{T_{eff}}}\;
\int d^Dx \left[ \frac12(\nabla\phi)^2 \; - \mu^2 \; e^{\phi({\vec
x})}\right]}\; , 
\end{equation}

where

\begin{equation}\label{paramD}
\mu^2 =  {{C_D} \over {(2\pi)^{D/2}}}\;
 z\; G \, m^{2+D/2} \, T^{D/2-1} 
\quad , \quad T_{eff} =  C_D \; {{G\; m^2}\over {T}} \; .
\end{equation}

We have then transformed
the partition function for the $D$-dimensional
gas of particles in gravitational interaction into the  partition
function for a 
scalar field $\phi$ with exponential interaction.  
The effective
temperature $ T_{eff} $ for the $\phi$-field partition function is
{\bf inversely} 
proportional to $ T $ for {\bf any} space dimension. The characteristic length
$\mu^{-1}$ behaves as $ \sim T^{-(D-2)/4} $.

\section{Scaling behaviour}

We derive here the scaling behaviour of the $\phi$ field following the
general renormalization group 
arguments in the theory of critical phenomena \cite{kgw,dg}

\subsection{Classical Scale Invariance}

Let us investigate how the  action (\ref{acci}) transforms under scale
transformations 
\begin{equation}\label{trafoS}
{\vec x} \to {\vec x}_{\lambda} \equiv \lambda{\vec x} \; ,
\end{equation}
where $\lambda$ is an arbitrary real number.

In $D$-dimensions the action takes the form
\begin{equation}\label{acciD}
S[\phi(.)] \equiv  {1\over{T_{eff}}}\;
\int d^D x \left[ \frac12(\nabla\phi)^2 \; - \mu^2 \; e^{\phi({\vec
x})}\right] \; .
\end{equation}

We define the scale transformed field $\phi_{\lambda}({\vec x})$ as follows
\begin{equation}\label{filam}
\phi_{\lambda}({\vec x}) \equiv \phi(\lambda{\vec x}) +\log\lambda^2
\; .
\end{equation}
Hence,
$$
(\nabla\phi_{\lambda}({\vec x}))^2 = \lambda^2 \; (\nabla_{x_{\lambda}}
\phi({\vec x}_{\lambda}))^2
\quad , \quad  e^{\phi_{\lambda}({\vec x})}=  \lambda^2  \;
 e^{\phi({\vec x}_{\lambda})}
$$
We find upon changing the integration variable in eq.(\ref{acciD})
from $ {\vec x} $  to  $ {\vec x}_{\lambda} $
\begin{equation}\label{covdil}
S[\phi_{\lambda}(.)] = \lambda^{2-D} \; S[\phi(.)] 
\end{equation}

We thus see that the  action (\ref{acciD})  {\bf scales} under dilatations
in spite of the 
fact that it contains the dimensionful parameter $ \mu^2 $. 
This remarkable scaling property is of course a consequence of the
scale   behaviour of the gravitational interaction  (\ref{hamiD}).

In particular, in $ D = 2 $ 
the action (\ref{acciD}) is scale invariant. In such
special case, it is moreover conformal invariant.

\bigskip

The (Noether) current associated to the scale transformations 
(\ref{trafoS}) is 
\begin{equation}
J_i({\vec x}) = x_j\; T_{ i j} ({\vec x}) + 2 \;\nabla_i\phi({\vec x})\; ,
\end{equation}
where $  T_{ij} ({\vec x}) $ is the stress tensor
$$
 T_{ i j} ({\vec x}) =  \nabla_i\phi({\vec x}) \; \nabla_j\phi({\vec x})
- \delta_{ij} \; L
$$
and $L \equiv \frac12(\nabla\phi)^2 \; - \mu^2 \; e^{\phi({\vec x})}
$ stands for the action density. That is,
$$
J_i({\vec x}) = ({\vec x}. \nabla\phi + 2)\;  \nabla_i\phi({\vec x})-
x_i \; \left[ \frac12(\nabla\phi)^2 \; - \mu^2 \; e^{\phi({\vec x})}\right]
$$
By using the classical equation of motion (\ref{eqMov}), we then find
$$
 \nabla_i J_i({\vec x}) = (2 - D) L \; .
$$
This non-zero divergence is due to the variation of the action under
dilatations [eq.  (\ref{covdil})].

\bigskip

If $\phi({\vec x})$ is a stationary point of the action (\ref{acciD}):
\begin{equation}\label{eqMov}
\nabla^2\phi({\vec x}) +  \mu^2 \; e^{\phi({\vec x})} = 0 \; ,
\end{equation}
then $ \phi_{\lambda}({\vec x}) $ [defined by eq.(\ref{filam})] is
also a stationary point:
$$
\nabla^2\phi_{\lambda}({\vec x}) +  \mu^2 \; e^{\phi_{\lambda}({\vec
x})} = 0 \; . 
$$

A rotationally invariant stationary point is given by
\begin{equation}\label{fic}
\phi^c(r) = \log{{2(D-2)}\over { \mu^2 r^2}} \; .
\end{equation}
This singular solution is {\bf invariant} under the scale
transformations (\ref{filam}). That is
$$
\phi^c_{\lambda}(r) =\phi^c(r) \; .
$$
Eq.(\ref{fic}) is dilatation and rotation invariant. 
It provides the {\bf most symmetric} stationary point of
the action. Notice that there are no constant stationary solutions
besides the singular solution $ \phi_0 = -\infty $.

The introduction of the short distance cutoff $a$, eq.(\ref{hamiR}), 
spoils the scale behaviour (\ref{covdil}). For the cutoff theory
from eqs.(\ref{Sregu}) and (\ref{trafoS})-(\ref{filam}),
we have instead
$$
S_a[\phi_{\lambda}(.)] = \lambda^{2-D} \; S_{\lambda a}[\phi(.)] \; .
$$

For $ r \sim a $,  eq.(\ref{fic}) does not hold anymore for the
spherically symmetric solution $\phi^c(r)$. For small $ r $ and $ a $,
using  eqs.(\ref{Sregu}-\ref{Sar}) we have
\begin{equation}\label{fica}
\phi^c(r)  \buildrel{r\to 0}\over = -{{ \mu^2 r^2}\over {2 D}} 
+ O(r^2, r^2 a^2)\; .
\end{equation}
That is, $\phi^c(r)$ is regular at $r = 0$ in the presence  of the
cutoff $a$.

\subsection{Thermal Fluctuations}

In this section we compute the partition function eqs.(\ref{zfi}) and
(\ref{zfiJ}) by saddle point methods.

Eq.(\ref{eqMov}) admits  only one constant stationary solution
\begin{equation}\label{fiS}
 \phi_0 = -\infty \; .
\end{equation}

In order to make such solution finite we now introduce a
regularization term $ \; \epsilon \, \mu^2 \, \phi({\vec x}) $ with $
\epsilon << 1 $ in the action  $ S $ [eq.(\ref{acci})]. This
corresponds to an action density
\begin{equation}\label{densac}
L =  \frac12(\nabla\phi)^2 \; + \; u(\phi)
\end{equation}
where
$$
 u(\phi) =  - \mu^2 \; e^{\phi({\vec x})} +  \epsilon \; \mu^2 \;
 \phi({\vec x}) \; .
$$
This extra term can be obtained by adding a small constant term $
-\epsilon \; \mu^2/T_{eff} $ to $\rho({\vec x})$ in eqs.(\ref{defro}) -
(\ref{reprf}). This is a simple way to make $ \phi_0 $ finite.

We get in this way a constant stationary point at $   \phi_0 =
\log\epsilon $ where $ u'(\phi_0) = 0 $. However, scale invariance is
broken since $  u''(\phi_0) = - \epsilon \; \mu^2 \neq 0 $. We can add
a second regularization term to   $ \; \frac12 \, \delta \, \mu^2 \;
 \phi({\vec x})^2 \; $  to $ L $, (with $  \delta << 1 $) in order to enforce $
u''(\phi_0) = 0 $. This quadratic term amounts to a long-range
shielding of the gravitational force. 
We finally set:
$$
 u(\phi) =  - \mu^2 \left[  e^{\phi({\vec x})} -  \epsilon \;
 \phi({\vec x}) -  \frac12 \; \delta \; \phi({\vec x})^2 \right]
 \; ,
$$
where the two regularization parameters $ \epsilon$ and $ \delta $ are
related by 
$$ 
 \epsilon( \delta ) = \delta [1 - \log \delta] \; ,
$$
and the stationary point has the value
$$
  \phi_0 =\log\delta \; .
$$

Expanding around $ \phi_0 $ 
$$
\phi({\vec x}) =  \phi_0 + g \; \chi({\vec x})
$$
where $ g \equiv \sqrt{\mu^{D-2} \, T_{eff}} $ and 
$ \chi({\vec x}) $ is the fluctuation field, yields
\begin{equation}\label{fiinfi}
\frac{1}{g^2}\;  L =  \frac12 \; (\nabla\chi)^2 \;
- {{\mu^2 \delta}\over {g^2}} \left[ e^{g \chi} -1 - g \; \chi - 
 \frac12 \; g^2 \;  \chi^2 \right]
\end{equation}
We see
perturbatively in $g$ that  $ \chi({\vec x}) $ is a {\bf massless} field.

\bigskip

Concerning the boundary conditions, we must consider the system inside
a large sphere of radius $R \; ( 10^{-4}\; - \; 10^{-2}
\;  pc \;   \leq     R   \leq 100\;  pc )$. That is, all
integrals are computed over such large sphere.

\bigskip

Using eq.(\ref{rouno}) the particle density takes now the form
$$
 \rho({\vec r}) =  -{1 \over {T_{eff}}}\;  \nabla^2 \phi({\vec r}) = 
 -{g \over {T_{eff}}}\;  \nabla^2 \chi({\vec r}) =  {{\mu^2
 \delta}\over { T_{eff} }} \left[ e^{g \chi({\vec r})} -1 - g
 \chi({\vec r})  \right]   \; .
$$
It is convenient to renormalize the  particle density 
by its stationary value $ \delta = e^{\phi_0} $,
\begin{equation}\label{renro}
 \rho({\vec r})_{ren} \equiv \frac{1}{\delta} \;  \rho({\vec r}) =
 {{\mu^D}\over {g^2 }} \left[ e^{g \chi({\vec r})} -1 - g \chi({\vec
 r})  \right]   \; .
\end{equation}
We see that in the $ \delta \to 0 $ limit the interaction  in 
eq.(\ref{fiinfi}) vanishes. No infrared divergences  appear in the
Feynman graphs calculations,  since we work on a
very large but finite volume of size $ R $. Hence, in the  $\delta \to
0 $ limit, the whole perturbation series around $ \phi_0 $  reduces  to the
zeroth order term.

The constant saddle point $ \phi_0 $ fails to catch up  the whole field
theory content. In fact, more information arises perturbing around the
stationary point $ \phi^c(r) $ given by eq.(\ref{fic}) \cite{fut}. 

Using eqs.(\ref{corr2}), (\ref{Dgreen}),  (\ref{fiinfi}) and
(\ref{renro}) we obtain for the density correlator in the   $ \delta
\to 0 $ limit, 
$$
C({\vec r_1},{\vec r_2}) =   {{\mu^{2 D}}\over {g^4}}\;
\left\{ \exp\!\left[{{g^2}\over { C_D\;  \, 
  \left( \mu \, | {\vec  r_1} - {\vec  r_2}|\right)^{D-2}}} \right] -1 
- {{g^2}\over {  C_D \;    \left(  \mu \, | {\vec  r_1} - {\vec
r_2}|\right)^{D-2}}} \right\} \; .
$$

For large distances, we find

\begin{equation}\label{corrasi}
 C({\vec r_1},{\vec r_2}) \buildrel{  | {\vec  r_1} - {\vec  r_2}|\to
\infty}\over =  {{ \mu^4 }\over {2\, C_D^2 \; 
 | {\vec  r_1} - {\vec  r_2}|^{2(D-2)}}} + O\left( \; | {\vec
r_1} - {\vec r_2}|^{-3(D-2)}\right)\; . 
\end{equation}

That is, the  $\phi$-field theory  {\bf scales}.  Namely, the
theory behaves  critically for a {\bf continuum set} of  values of $\mu$ and
$ T_{eff} $.  

Notice that the
density correlator $C({\vec r_1},{\vec r_2})$ behaves for large
distances as the correlator of $ \chi({\vec r})^2 $. This stems from
the fact that $ \chi({\vec r})^2 $ is the most relevant operator in
the series expansion of the density (\ref{renro})
\begin{equation}\label{denser}
 \rho({\vec r})_{ren} = \frac12 \; \mu^D \;  \chi({\vec r})^2 + O(\chi^3)\; .
\end{equation}

As remarked above, the constant stationary point $ \phi_0 = \log\delta \to
-\infty $ only produces  the zeroth order of perturbation theory.
More information arises perturbing around the stationary point $ \phi^c(r) $
given by eq.(\ref{fic}) \cite{fut}. 

\subsection{Renormalization Group  Finite Size Scaling Analysis}

As is well known \cite{kgw,dg,nn}, physical quantities  for {\bf
infinite} volume systems diverge at the critical point  as $ \Lambda $ to a
negative power.  $ \Lambda $ measures the distance to the critical
point. (In condensed matter and spin systems, $ \Lambda $ is
proportional to the temperature minus the critical temperature \cite{dg,nn}).
One has  for the correlation length  $ \xi $, 
$$ 
\xi( \Lambda ) \sim  \Lambda^{-\nu} \; ,
$$
and for the specific heat (per unit volume) $ {\cal C} $,
\begin{equation}\label{calor}
 {\cal C} \sim  \Lambda^{-\alpha}  \; .
\end{equation}
Correlation functions scale at criticality. For example,  the 
scalar field $\phi$ (which in spin systems describes the magnetization)
scales as,
$$
<\phi({\vec r})\phi(0)> \sim r^{-1-\eta} \; .
$$
The critical exponents $\nu, \;\alpha $ and $ \eta $ are pure numbers
that depend only on the universality class  \cite{kgw,dg,nn}.

For a {\bf finite} volume system, all physical  quantities  are  {\bf
finite} at the critical point. Indeed, for a  system whose  size $ R $
is large, the  physical  magnitudes
take large values at the critical point. Thus, for large   $ R $, one can
use the infinite volume theory to treat finite size systems at
criticality. In particular,  the correlation length provides the
relevant physical length $ \xi \sim R $. This implies that
\begin{equation}\label{fss}
\Lambda \sim R^{-1/\nu} \; .
\end{equation}
We can apply these concepts to the  $\phi$-theory 
since, as we have seen in the previous section, it 
exhibits scaling in a finite volume $\sim R^3 $. 
Namely, the two points correlation function exhibits a power-like
behaviour in perturbation theory as shown by  eq.(\ref{corrasi}). This happens 
 for a  {\bf continuum set} of  values of 
$T_{eff}$ and $\mu^2$. Therefore, changing $\mu^2/T_{eff}$ keeps the
theory in the scaling region. 
At the point $ \mu^2/T_{eff} = 0 $, the partition function $ {\cal Z}
$ is singular. From eq.(\ref{muyT}),  we shall thus identify
\begin{equation}\label{zcritico}
  \Lambda \equiv   {{\mu^2}\over{T_{eff}}} = z\,
  \left({{mT}\over{2\pi}}\right)^{3/2} \; .
\end{equation}

Notice that the critical point $  \Lambda = 0 $, corresponds to zero
fugacity. 

Thus,  the partition function in the scaling regime can be written as 
\begin{equation}\label{Zsca1}
{\cal Z}(\Lambda) = 
 \int\int\;  {\cal D}\phi\;  e^{ -S^* + \Lambda
\int d^Dx  \; e^{\phi({\vec x})}\;}\; ,
\end{equation}
where $S^*$ stands for the action (\ref{acci}) at the critical point
$\Lambda = 0 $.

We define the renormalized mass  density  as
\begin{equation}\label {dfensi}
m\, \rho({\vec x})_{ren} \equiv m\, \,  e^{\phi({\vec x})}
\end{equation}
and we identify it with the  energy density in the renormalization
group. [Also called the `thermal perturbation operator'].
This identification  follows from the fact that they are the most
relevant positive definite operators. Moreover, such  identification is
supported by the perturbative result (\ref{denser}).

In the scaling regime we have \cite{dg} for the logarithm of the
partition function
\begin{equation}\label{Zsca2}
{1 \over V} \; \log{\cal Z}(\Lambda) = {K \over{(2-\alpha)(1-\alpha)}}\;
\Lambda^{2-\alpha} + F(\Lambda) \; ,
\end{equation}
where   $  F(\Lambda) $ is an analytic
function of $ \Lambda $ around the origin 
$$ 
F(\Lambda) = F_0 + a \; \Lambda + \frac12 \, b  \; \Lambda^2 + \ldots \; .
$$ 
 $ V = R^D $ stands for the volume and $ F_0, \; K, \; a $ and $ b $
are constants. 

Calculating the logarithmic derivative of ${\cal Z}(\Lambda)$ with
respect to $ \Lambda $ from eqs.(\ref{Zsca1}) and from (\ref{Zsca2})
and equating the results yields
\begin{equation}\label{masaR}
{1 \over V} \;{{\partial}\over{\partial\Lambda}}\log{\cal Z}(\Lambda)=
a +  {K \over{1-\alpha}}\,
\Lambda^{1-\alpha}
+ \ldots = {1 \over V} \int d^Dx  \; <e^{\phi({\vec x})}>\; .
\end{equation}
where we used  the scaling
relation $ \alpha = 2 - \nu D $ \cite{dg,nn}.

We can apply here finite size scaling arguments and 
 replace $\Lambda$ by $\sim R^{-\frac{1}{\nu}}$ [eq.(\ref{fss})],
$$
{{\partial}\over{\partial\Lambda}}\log{\cal Z}(\Lambda)= V \, a +
 {K \over{1-\alpha}}\, R^{1/\nu} + \ldots\; . 
$$

Recalling eq.(\ref{dfensi}), we can express the mass contained in a region
of size $ R $ as
\begin{equation}\label{defM}
M(R) = m  \int^R e^{\phi({\vec x})} \; d^Dx  \; .
\end{equation}
Using  eq.(\ref{masaR}) we find
$$
<M(R)> =  m  \, V \, a +m  \, {K \over{1-\alpha}}\; R^{ \frac1{\nu}} +
\ldots\; . 
$$
and
\begin{equation}\label{Sdensi}
<\rho({\vec r})> = m  a \; +m \, {K
\over{\nu(1-\alpha)\Omega_D}}\;r^{ \frac1{\nu}-D} \quad {\rm for}\; r
\;  {\rm of ~ order}\;\sim R .
\end{equation}
where $ \Omega_D $ is the surface of the unit sphere in $D$-dimensions.


The energy density correlation function is known in general in the
scaling region (see  refs.\cite{dg} -\cite{nn}).
We can therefore write for the density-density correlators
(\ref{corre}) in $ D $ space dimensions
\begin{equation}\label{corrG}
C({\vec r_1},{\vec r_2})\sim |{\vec r_1} -{\vec r_2}|^{\frac2{\nu} -2D} \; .
\end{equation}
where both $ {\vec r_1} $ and $ {\vec r_2} $ are  inside the
finite  volume $ \sim R^D $.

The perturbative calculation (\ref{corrasi})  matches
with this result for $ \nu = \frac12 $. That is, the mean field
value for the exponent $ \nu $. 

Let us now compute the second derivative of $ \log{\cal Z}(\Lambda) $
with respect to $\Lambda$ in two ways.  We find from eq.(\ref{Zsca2}) 
$$
{{\partial^2}\over{\partial\Lambda^2}}\log{\cal Z}(\Lambda)= V\left[
\Lambda^{-\alpha} \, K + b  + \ldots  \right] \; .
$$
We get from eq.(\ref{Zsca1}),
\begin{equation}\label{flucM}
{{\partial^2}\over{\partial\Lambda^2}}\log{\cal Z}(\Lambda)=
\int d^Dx\; d^Dy\; C({\vec x},{\vec y}) \sim R^D \int^R 
{{ d^3x}\over{x^{2D - 2d_H}}}  \sim \Lambda^{-2}\sim R^D \;
\Lambda^{-\alpha} 
\end{equation}
where we used eq.(\ref{fss}), eq.(\ref{corrG}) and the scaling
relation $ \alpha = 2 - \nu D $ \cite{dg,nn}. We
conclude that the scaling 
behaviours,   eq.(\ref{Zsca2})  for the partition function, 
eq.(\ref{calor}) for the specific heat and  eq.(\ref{corrG}) for the
two points correlator are consistent.
In addition,   eqs.(\ref{defM}) and (\ref{flucM})
yield for the  mass fluctuations  squared
$$
(\Delta M(R))^2 \equiv  \; <M^2> -<M>^2  \; \sim
\int d^Dx\; d^Dy\; C({\vec x},{\vec y}) \sim R^{2d_H}\; .
$$
Hence, 
\begin{equation}\label{Msca}
\Delta M(R)  \sim  R^{d_H}\; .
\end{equation}

\bigskip

The scaling exponent $\nu$ can be identified with the inverse
Haussdorf (fractal) dimension $d_H$ of the system
$$
d_H = \frac1{\nu} \; .
$$
In this way, $ \Delta M \sim R^{d_H} $ according to the usual
definition of fractal dimensions \cite{sta}.

\medskip

Using eq.(\ref{corrG})  we can compute the average potential  energy
in three space dimensions as
$$
< {\cal V}> =   \frac12 \,\beta \, G \, m^2
\int_{ | {\vec x} - {\vec y}|> a }^R \;
{{d^3x\, d^3y}\over { | {\vec x} - {\vec y}|}}\;  C({\vec x},{\vec y})
\sim R^{\frac2{\nu} -1} \; .
$$

From here and eq.(\ref{Msca}) we get as  virial estimate for the atoms
kinetic energy 
$$
<v^2> = {{< {\cal V}>}\over {< \Delta M(R)>}} \sim  R^{\frac1{\nu} -1} \; .
$$
This corresponds to a velocity dispersion
\begin{equation}\label{Vsca}
\Delta v \sim R^{\frac12(\frac1{\nu} -1)} \; .
\end{equation}
That is,  we predict [see eq.(\ref{vobser})] a new scaling relation
$$
q =\frac12\left(\frac1{\nu} -1\right) =\frac12(d_H -1)  \; .
$$

\bigskip

The calculation of the critical amplitudes [that is, the coefficients in 
front of the powers of $ R $ in eqs.(\ref{corrG}), (\ref{Msca}) and
 (\ref{Vsca})] is beyond the scope of the present paper \cite{fut}.

\subsection{Values of the scaling exponents and the fractal dimensions}

The scaling exponents $ \nu , \; \alpha $ considered in sec IIIC can
be computed through the renormalization group approach. The case of a 
 {\bf single} component scalar field has been extensively studied 
in the literature \cite{dg,nn,grexa}. Very probably, there is an
unique, infrared stable fixed point in three space dimensions: the
Ising  model fixed point. Such  non-perturbative fixed point is
reached in the long scale regime independently of the initial shape of
the interaction $ u(\phi) $ [eq.(\ref{densac})]  \cite{grexa}.

The numerical values of the scaling exponents associated to the 
Ising  model fixed point are

\begin{equation}\label{Ising}
\nu = 0.631\ldots   \quad , \quad d_H = 1.585\ldots   \quad , \quad \eta =
0.037\ldots  \quad {\rm and} \quad \alpha = 0.107\ldots  \; \; .
\end{equation}

\bigskip

In the $\phi$ field model there are two dimensionful parameters: $\mu$
and $T_{eff}$. The dimensionless combination 
\begin{equation}\label{defg}
g^2 = \mu \, T_{eff} =  (8 \pi)^{3/4}\; \sqrt{z} \; \; {{G^{3/2}\;
m^{15/4}}\over T^{3/4}} 
\end{equation}
acts as the coupling constant for the non-linear fluctuations of the
field $\phi$.  

Let us consider a gas formed by neutral hydrogen at thermal equilibrium with
the cosmic microwave background. We set $ T = 2.73\, K $ and  estimate
the fugacity $ z $ using the ideal gas value
$$
z = \left( {{2\pi}\over {m T } }\right)^{3/2}\; \rho \; .
$$
Here we use $  \rho = \delta_0 $ atoms cm$^{-3}$ for the ISM density and
$  \delta_0 \simeq 10^{10} $. Eq. eqs.(\ref{muyT}) yields
\begin{equation}\label{valN}
{1 \over {\mu }} = 2.7 \; {1 \over { \sqrt{\delta_0}}}\; {\rm  AU} \sim
30 \; {\rm AU} \quad {\rm and} \quad 
g^2 = \mu \, T_{eff} = 4.9 \; 10^{-58} \; \sqrt{\delta_0} \sim 5 \,10^{-53}
 \; . 
\end{equation}

This extremely low value for $g^2 $ 
suggests that the perturbative calculation [sec. IIIB] may apply here 
yielding the mean field values for the exponents, i. e.  
\begin{equation}\label{campM}
\nu = 1/2 \quad ,  \quad d_H = 2   \quad , \quad \eta =0 \quad {\rm
and } \quad \alpha = 0 \; .
\end{equation}
That is, the effective coupling constant grows with
the scale according to the renormalization group flow (towards the
Ising fixed point). Now, if the extremely low value of the initial
coupling eq.(\ref{valN}) applies, the perturbative result (mean field)
will hold for many scales (the effective $ g $ grows roughly as the
length).

$ \mu^{-1} $ indicates the order of the smallest distance where the
 scaling regime applies.  A safe
lower bound supported by observations is around $20$ AU $\sim 3.\,
10^{14}$ cm , in agreement with our estimate.

\bigskip

Our theoretical predictions for $ \Delta M(R) $ and $ \Delta v $
[eqs.(\ref{Msca}) and (\ref{Vsca})] both for 
the Ising  eq.(\ref{Ising}) and for the mean field values
eq.(\ref{campM}), are in  agreement with the astronomical
observations [eq.(\ref{vobser})]. The present observational bounds on
the data are  larger than the difference between the mean field and
Ising values of  the exponents $ d_H $ and $ q $. 

Further theoretical work in the $\phi$-theory will determine whether
the scaling behaviour is given by the mean field or by the Ising fixed
point \cite{fut}.

\subsection{The two dimensional gas and random surfaces fractal dimensions} 

In the two dimensional case ($D=2$) the partition function
(\ref{zfiD}) describes  the Liouville model that arises in string
theory\cite{poly} and in the theory of random surfaces 
(also called two-dimensional quantum gravity).
For strings in $c$-dimensional Euclidean space the  partition function
takes the form\cite{poly}
\begin{equation}\label{zfiL}
{\cal Z}_c =  \int\int\;  {\cal D}\phi\;  e^{ -{{26-c}\over{24\pi}}\;
\int d^2x \left[ \frac12(\nabla\phi)^2 \; + \mu^2 \; e^{\phi({\vec
x})}\right]}\; . 
\end{equation}
This coincides with eq.(\ref{zfiD}) at $D=2$ provided we flip the sign
of $ \mu^2 $ and identify the parameters (\ref{paramD}) as follows,
\begin{equation}
T = G m^2\; {{26-c}\over 12}\quad , \quad \mu^2 = z G m^3 \; .
\end{equation}

Ref.\cite{amb} states that $ d_H = 4 $ for $ c \leq 1 $, $ d_H = 3 $ for
$ c = 2 $ and  $d_H = 2$ for
$ c \geq 4 $. In our context this means
$$
d_H =2 \;\; {\rm for}~~ T \leq    \frac{25}{12} \;  G m^2 \quad ,    \quad
 d_H = 3 \;\; {\rm for}~~ T = 2\, G m^2  \quad
{\rm and} \quad
 d_H =4 \;\; {\rm for}~~ T \geq \frac{11}6  \;  G m^2 \; .
$$

For $ c \to \infty , \; g^2 \to 0 $ and we can use the perturbative
result (\ref{corrasi}) yielding $ \nu = \frac12 , \; d_H = 2 $
in agreement with the above discussion for $ c \geq 4 $. 

\subsection{Stationary points and the Jeans length}

The stationary points of the $\phi$-field partition function (\ref{zfi}) 
are given by the non-linear partial differential equation
$$
\nabla^2\phi = -\mu^2\,  e^{\phi({\vec x})} \; .
$$
In terms of the gravitational potential $U({\vec x})$ [see eq. (\ref{Ufi})],
this takes the form

\begin{equation}\label{equih}
\nabla^2U({\vec r}) = 4 \pi G \, z \,  m
\left({{mT}\over{2\pi}}\right)^{3/2} \,  e^{ - \frac{m}{T}\,U({\vec r})} \; .
\end{equation}
This corresponds to the Poisson equation for a thermal matter distribution
fulfilling an ideal gas in hydrostatic equilibrium,
as can be seen as follows \cite{sas}.
The hydrostatic equilibrium condition 
$$
\nabla P({\vec r}) = - m \, \rho({\vec r}) \; \nabla U({\vec r})\; ,
$$
where $ P({\vec r}) $ stands for the pressure, combined with the
equation of state for the ideal gas
$$
P = T \rho \; ,
$$
yields for the particle density
$$
 \rho({\vec r}) =  \rho_0 \; e^{ - \frac{m}{T}\,U({\vec r})} \; ,
$$
where $ \rho_0 $ is a constant. Inserting this relation into the
Poisson equation 
$$
\nabla^2U({\vec r}) = 4 \pi G\, m \, \rho({\vec r})
$$
yields eq.(\ref{equih}) with 
\begin{equation} \label{RO0}
  \rho_0 =  z \,\left({{mT}\over{2\pi}}
\right)^{3/2} \; . 
\end{equation}

For large $ r $,  eq.(\ref{equih}) gives a density decaying as $
r^{-2} $ ,
\begin{equation}
 \rho({\vec r}) \buildrel{r\to \infty}\over = {T\over{2\pi G m}}\,
\frac1{r^2} \,\left[ 1 + O\left(\frac1{\sqrt{r}} \right)  \right]
\quad , \quad U({\vec r}) \buildrel{r\to \infty}\over = 
\frac{T}{m}\;\log\left[{{2\pi G  \rho_0}\over T}\; r^2\right] +
O\left(\frac1{\sqrt{r}} \right) \; .
\end{equation}
Notice that this density, which describes a single stationary solution, 
decays for large $r$ {\bf faster} than the density (\ref{Sdensi}) governed by
thermal fluctuations.

\bigskip

Spherically symmetric solutions of eq.(\ref{equih}) has been studied 
in detail \cite{chandra}.
The small fluctuations around such isothermal spherical solutions
as well as the stability problem were studied in \cite{kh}.

\bigskip

The Jeans distance is in this context,
\begin{equation}\label{distaJ}
d_J \equiv \sqrt{ 3 T \over m}\; {1 \over{\sqrt{G\, m \, \rho_0}}} = 
{{ \sqrt{ 3}\; (2\pi)^{3/4}}\over{  \sqrt{z\, G}\; m^{7/4}\; T^{1/4}}}
\; .
\end{equation}
This distance precisely coincides with $ \mu^{-1} $ [see eq.(\ref{muyT})] up to
an inessential numerical coefficient ($\sqrt{12/\pi}$). Hence,  $
\mu $, the only dimensionful parameter in the $\phi$-theory can
be interpreted as the inverse of the Jeans distance.

We want to notice that in the critical regime,  $ d_J $ grows as
\begin{equation}\label{escaA}
d_J \sim R^{d_H/2} \; ,
\end{equation}
since $  \rho_0 = \Lambda \sim R^{-d_H} $ vanishes as 
can be seen from  eqs.(\ref{fss}), (\ref{zcritico}) and
(\ref{distaJ}). In this tree level estimate  we should use for consistency
the mean field value $ d_H = 2 $, which yields  $ d_J \sim R$.

This  shows that the Jeans distance is  of the  order of the {\bf size} of the
system. The  Jeans distance   {\bf scales} and the instability is
therefore present for all sizes $ R $.

Had  $ d_J $ being   of order larger than $ R $, the Jeans instability
would be absent. 

The fact that  the Jeans instability is  present  {\bf precisely}
at $  d_J \sim R $ is probably essential to  the scaling regime and to
the self-similar (fractal) structure of the gravitational gas. 

The dimensionless coupling constant $ g^2 $ can be written from
eqs.(\ref{zcritico}) and (\ref{defg}) as 
$$
g^2 = \left( 2m \sqrt{{\pi \, G}\over T}\right)^3 \sqrt{\Lambda}\; .
$$
Hence, the tree level coupling scales as
$$
g^2 \sim R^{-1} \; .
$$
Direct perturbative calculations explicitly exhibit such scaling behaviour
\cite{fut}. 

We can express $ g^2 $ in terms of $ d_J $ and $ \rho_0 $ as follows,
$$
g^2 = {{(12 \pi)^{3/2}}\over { \rho_0 \;   d_J^3}} = {{\pi^2
\mu^3}\over { \rho_0}}
\; .
$$
This shows that  $ g^2 $ is, at the tree level, the inverse of the
number of particles inside a Jeans volume. 

Eq.(\ref{escaA}) applies to the tree level Jeans length or tree level 
 $ \mu^{-1} $. We can furthermore estimate the Jeans length using the
 renormalization group behaviour of the physical quantities derived in
 sec. III.C. Setting,
$$
<d_J> = {{<\Delta v>}\over {  \sqrt{G\,  m\,  <\Delta \rho>}}} \; ,
$$
we find from eqs.(\ref{Sdensi}) and (\ref{Vsca}),
$$
<d_J> \sim R \; .
$$
Namely, we find again that the Jeans length grows as the size $ R $.

\section{Discussion}

In previous sections we ignored gravitational forces external to the
gas like stars etc. Adding a  fixed external mass density $
\rho_{ext}({\vec r}) $ amounts to introduce an external source
$$
J({\vec r}) =  - T_{eff}\; \rho_{ext}({\vec r})\; ,
$$
in eq.(\ref{zfiJ}). Such term will obviously affect correlation
functions, the  mass density, etc. except when we look at the scaling
behaviour which is governed by the critical point. 
That is, the values we find for the scaling exponents $ d_H $ and $ q
$ are {\bf stable} under external perturbations.

\bigskip

We considered all atoms with the same mass in the gravitational gas.
It is easy to generalize the transformation into the $\phi$-field
presented in section II for a mixture of several kinds of atoms. Let
us consider $ n $ species of atoms with 
masses $ m_a, \; 1 \leq a \leq n $. Repeating the steps from
eq.(\ref{gfp}) to (\ref{acci})
yields again a field theory with a single scalar field  but the
action now takes the form
\begin{equation}\label{gasM}
S[\phi(.)] \equiv  {1\over{T_{eff}}}\;
\int d^3x \left[ \frac12(\nabla\phi)^2 \; - \sum_{a=1}^n \; \mu_a^2 \;
e^{{{m_a}\over m}\, \phi({\vec x})}\right] \; ,
\end{equation}
where
$$
\mu_a^2 = \sqrt{2\over {\pi}}\; z_a \; G \, m_a^{3/2} \, m^2 \, \sqrt{T}
\; ,
$$
and $ m $ is just a reference mass. 

Correlation functions, mass densities and other observables will 
obviously depend on the number of species, their masses and fugacities
but it is easy to see that the fixed points and scaling exponents are
exactly the {\bf same} as for the $\phi$-field theory (\ref{zfi})-(\ref{muyT}).

\bigskip

We want to notice that there is an important difference between the
behaviour of the gravitational gas and the spin models (and all other
statistical models in the same universality class). For the
gravitational gas we find scaling behaviour for a {\bf full range} of
temperatures and couplings. For spin models scaling only appears 
at the critical value of the temperature. At the critical temperature
 the correlation length $ \xi $ is infinite and the theory is
massless. 
For temperatures near the critical one,  i. e. in the critical
domain,  $ \xi $ is finite (although very
large compared with the lattice spacing) and the correlation functions
decrease as $ \sim e^{ - r/\xi} $ for large distances $ r $.  
Fluctuations of the relevant operators support perturbations which can
be interpreted as massive excitations. Such
(massive) behaviour does not appear for the gravitational gas. The ISM
correlators scale exhibiting power-law behaviour. This feature is
connected with the scale invariant character of the Newtonian force
and its infinite range.

\bigskip

The hypothesis of strict thermal equilibrium does not apply to the ISM as 
a whole where temperatures range from $ 5 $ to $ 50 $ K and even $ 1000 $ K. 
However, since the scaling behaviour is independent of the temperature,
it applies to {\bf each} region of the ISM in thermal equilibrium.
Therefore, our theory applies provided thermal equilibrium holds
in regions or clouds. 

 We have developed here the theory of a gravitationally interacting
ensemble of bodies at a  fixed temperature.  In a real situation like the ISM, 
 gravitational perturbations  from external masses,
as well as other perturbations are present.
We have shown that the scaling solution is stable
with respect to the gravitational perturbations. It is well known that 
solutions based on a fixed point are generally quite robust.

Our theory  especially applies  to the interstellar medium far from 
star forming regions, which can be locally far from thermal equilibrium,
and where ionised gas at 10$^4$K together with coronal gas at 10$^6$K 
can coexist with the cold interstellar medium. In the outer parts of
galaxies, devoid of star formation, the ideal isothermal conditions 
are met \cite{pcm}. Inside the Galaxy, large regions satisfy also
the near isothermal criterium, and these are precisely the regions
where scaling 
laws are the best verified. Globally over the Galaxy, the fraction
of the gas in the hot ionised phase represents a negligible mass, 
a few percents, although occupying a significant volume. Hence, this
hot ionised gas is a perturbation which may not  change the fixed point
behaviour of the thermal  self-gravitating gas.

\bigskip

In ref.\cite{pm} a connection between a gravitational gas of galaxies 
in an expanding universe and the
Ising model is conjectured. However, the unproven identification made  
in ref.\cite{pm} of the  mass density contrast with the Ising spin leads to
scaling exponents  different from ours.

\acknowledgements
H J de V and N S  thank D. Boyanovsky and M. D'Attanasio for discussions.


\begin{thebibliography}{99} 
\bibitem{larson}
R. B. Larson, M.N.R.A.S. {\bf 194}, 809 (1981)
\bibitem{obser}
J. M. Scalo, in `Interstellar Processes', D.J. Hollenbach and 

H.A. Thronson Eds., D. Reidel Pub. Co, p. 349 (1987).

\bibitem{lar2} R. B. Larson, M.N.R.A.S. {\bf 256}, 641 (1992)

\bibitem{pcm} D. Pfenniger, F. Combes, L. Martinet, A\&A {\bf 285}, 79 (1994)

D. Pfenniger, F. Combes,  A\&A {\bf 285}, 94 (1994)  

\bibitem{llflu}
L. Landau and E. Lifchitz, M\'ecanique des Fluides, Eds. MIR, Moscou 1971.
\bibitem{weisz}
C.F. von Weizs\"acker, ApJ, {\bf 114}, 165 (1951).
\bibitem{leo}
L. P. Kadanoff, `From Order to Chaos', World. Sc. Pub.(1993).

\bibitem{origen}
S. Edward and A. Lenard, J. M. P. {\bf 3}, 778 (1962).

S. Albeverio and R. H{\o}egh-Krohn, C. M. P. {\bf 30}, 171 (1973).

\bibitem{stra} R. L. Stratonovich, Doklady, {\bf 2}, 146 (1958).

J. Hubbard, Phys. Rev. Lett, {\bf 3}, 77 (1959).

J. Zittartz, Z. Phys., {\bf 180}, 219 (1964).

\bibitem{ll} L. D. Landau and E. M. Lifchitz, Physique Statistique,
 4\`eme \'edition, Mir-Ellipses, 1996.

\bibitem{sam} S. Samuel, Phys. Rev. {\bf D 18}, 1916 (1978).

\bibitem{kgw}  K. G. Wilson,  Rev.\ Mod.\ Phys. {\bf 47}, 773 (1975)
  and  Rev.\ Mod.\ Phys.  {\bf 55}, 583 (1983).

\bibitem{dg} Phase transitions and Critical Phenomena vol. 6,

C. Domb \& M. S. Green, Academic Press, 1976.

\bibitem{nn} J. J. Binney, N. J. Dowrick, A. J. Fisher and 
 M. E. J. Newman,    

The Theory of Critical Phenomena,  Oxford Science Publication.

\bibitem{sta} See for example, 

H. Stanley in Fractals and Disordered Systems, 

A. Bunde and S. Havlin editors, Springer Verlag, 1991. 
\bibitem{sas} See for example, W. C. Saslaw, 
`Gravitational Physics of stellar and galactic systems',
Cambridge Univ. Press, 1987.

\bibitem{grexa} A. Hasenfratz and P.  Hasenfratz, Nucl.Phys. {\bf
B270}, 687 (1986).

T. R. Morris, Phys. Lett. {\bf B329}, 241 (1994) and {\bf B334}, 355 (1994).

\bibitem{chandra} S. Chandrasekhar, `An Introduction to
the Study of Stellar Structure',  

Chicago Univ. Press, 1939.

\bibitem{kh} G. Horwitz and J. Katz, 
Ap. J. {\bf 222}, 941 (1978) and  {\bf 223}, 311  (1978).

 J. Katz,  G. Horwitz and A. Dekel,
Ap. J. {\bf 223}, 299  (1978).

\bibitem{poly} A. M. Polyakov, Phys. Lett. {\bf B103}, 207 (1981).

\bibitem{amb} J. Ambj{\o}rn and  Y. Watabiki, Nucl. Phys. {\bf B 445},
129 (1995). 

 J. Ambj{\o}rn, J. Jurkiewicz  and  Y. Watabiki, Nucl. Phys. {\bf B 454},
313 (1995). 

Y. Watabiki, hep-th/9605185.
\bibitem{fut} H. J. de Vega,
 N. S\'anchez,  B. Semelin and F. Combes, in preparation.

\bibitem{klein} S.C. Kleiner, R.L. Dickman,
 
ApJ {\bf 286}, 255 (1984),
ApJ {\bf 295}, 466 (1985),
ApJ {\bf 312}, 837 (1987)


\bibitem{pm} J. P\'erez Mercader, T. Goldman, D. Hochberg and
R. Laflamme, 

astro-ph/9506127 and LAEFF-96/06. 

\end{thebibliography}
\end{document}